\documentclass[11pt]{article}
\usepackage{geometry}
\usepackage{makeidx}
\usepackage[T1]{fontenc}
\usepackage[dvipsnames,svgnames,table]{xcolor}
\usepackage{epstopdf}
\usepackage{epsfig}
\usepackage{textcomp}
\usepackage{hyperref}
\usepackage{amsmath}
\usepackage{amssymb}
\usepackage{multirow}
\usepackage{multicol}
\usepackage{booktabs}
\usepackage{lastpage}
\usepackage{hyperref} 
\usepackage{graphicx}
\usepackage{enumerate}
\usepackage{threeparttable}
\usepackage{float}
\usepackage{array}
\usepackage{cite}
\usepackage{color,soul}
\usepackage{listings}
\makeatletter
\renewcommand\section{\@startsection{section}{1}{\z@}%
                    {-2.5ex \@plus -1ex \@minus -.2ex}%
                    {2.3ex \@plus.2ex}%
                    {\normalfont\large\bfseries}}
\renewcommand\subsection{\@startsection{subsection}{1}{\z@}%
                    {-2.5ex \@plus -1ex \@minus -.2ex}%
                    {2.3ex \@plus.2ex}%
                    {\small\bfseries}}

\begin{document}
\bigskip
\title{\vskip -2.5 cm\textbf{DOME: Discrete oriented muon emission in GEANT4 simulations}}
\medskip
\author{\small A. Ilker Topuz$^{1,2}$, Madis Kiisk$^{1,3}$, Andrea Giammanco$^{2}$}
\medskip
\date{\small$^1$Institute of Physics, University of Tartu, W. Ostwaldi 1, 50411, Tartu, Estonia\\
$^2$Centre for Cosmology, Particle Physics and Phenomenology, Universit\'e catholique de Louvain, Chemin du Cyclotron 2, B-1348 Louvain-la-Neuve, Belgium\\
$^3$GScan OU, Maealuse 2/1, 12618 Tallinn, Estonia}
\maketitle
\begin{abstract}
Amongst various applications that experience a multi-directional particle source is the muon scattering tomography where a number of horizontal detectors of a limited angular acceptance conventionally track the cosmic-ray muons. In this study, we exhibit an elementary strategy that might be at disposal in diverse computational applications in the GEANT4 simulations with the purpose of hemispherical particle sources. To further detail, we initially generate random points on a spherical surface for a sphere of a practical radius by employing Gaussian distributions for the three components of the Cartesian coordinates, thereby obtaining a generating surface for the initial positions of the corresponding particles. Since we do not require the half bottom part of the produced spherical surface for our tomographic applications, we take the absolute value of the vertical component in the Cartesian coordinates by leading to a half-spherical shell, which is traditionally called a hemisphere. Last but not least, we direct the generated particles into the target material to be irradiated by favoring a selective momentum direction that is based on the vector construction between the random point on the hemispherical surface and the origin of the target material, hereby optimizing the particle loss through the source biasing. We also show a second scheme where the coordinate transformation is performed between the spherical coordinates and the Cartesian coordinates, and the above-mentioned procedure is applied to orient the generated muons towards the target material. In the end, a recipe hinged on the restrictive planes from our previous study is furthermore provided, and  we incorporate our strategies by using G4ParticleGun in the GEANT4 code. While we plan to exert our strategy in the computational practices for muon scattering tomography, these source schemes might find its straightforward applications in different neighboring fields including but not limited to atmospheric sciences, space engineering, and astrophysics where a 3D particle source is a necessity for the modeling goals.
\end{abstract}
\section{Introduction}
In practice, various shapes of radiation sources including but not limited to planar surfaces and parabolic beams have been utilized to mimic the associated reality in the desired applications, and one of these geometries includes hemispherical surfaces~\cite{pagano2021ecomug}. In this study, we describe the implementation steps of two schemes by aiming at building a hemispherical muon source where the generated particles are oriented towards a specific point or plane that we call selective momentum direction. While there exist different schemes to generate the 2D/3D sources, we prefer to use the existing algorithms in GEANT4~\cite{agostinelli2003geant4}, i.e. $\rm G4RandGauss::shoot()$ and $\rm G4UniformRand()$ as a distribution function. Whereas the geometrical shape of the 2D/3D sources plays an important role or is a parameter for this aim, the momentum direction is another variable that awaits for a user decision. In this study, we first generate a spherical surface by using three Gaussian distributions for the three components of the Cartesian coordinates and we direct the generated particles from their initial positions on this spherical surface to the preferred location(s) by using a vector construction as described in our previous study~\cite{topuz2022particle}. This methodology is called discrete oriented muon emission (DOME) where the kinetic energy of the generated particles is intentionally discrete for the computational purposes as already implemented in another study~\cite{topuz2022towards}. In the latter scheme, we generate the initial positions by randomizing the spherical variables, i.e. altitude and longitude, and we perform the coordinate transformation from the spherical coordinates to the Cartesian coordinates~\cite{marsaglia1972choosing, tashiro1977methods, weisstein2011disk}. We repeat the same operations as performed in the first scheme. This study is organized as follows. Section 2.1 describes the first scheme that is hinged on the Gaussian distribution functions, while section 2.2 consists of the second methodology founded on the coordinate transformation from the spherical coordinates to the Cartesian coordinates. Whereas an alternative focusing scheme is explained in section 3, we draw our conclusions in section 4.
\section{Central focus scheme}
\subsection{Generation through Gaussian distributions}
\begin{figure}[H]
\begin{center}
\includegraphics[width=12cm]{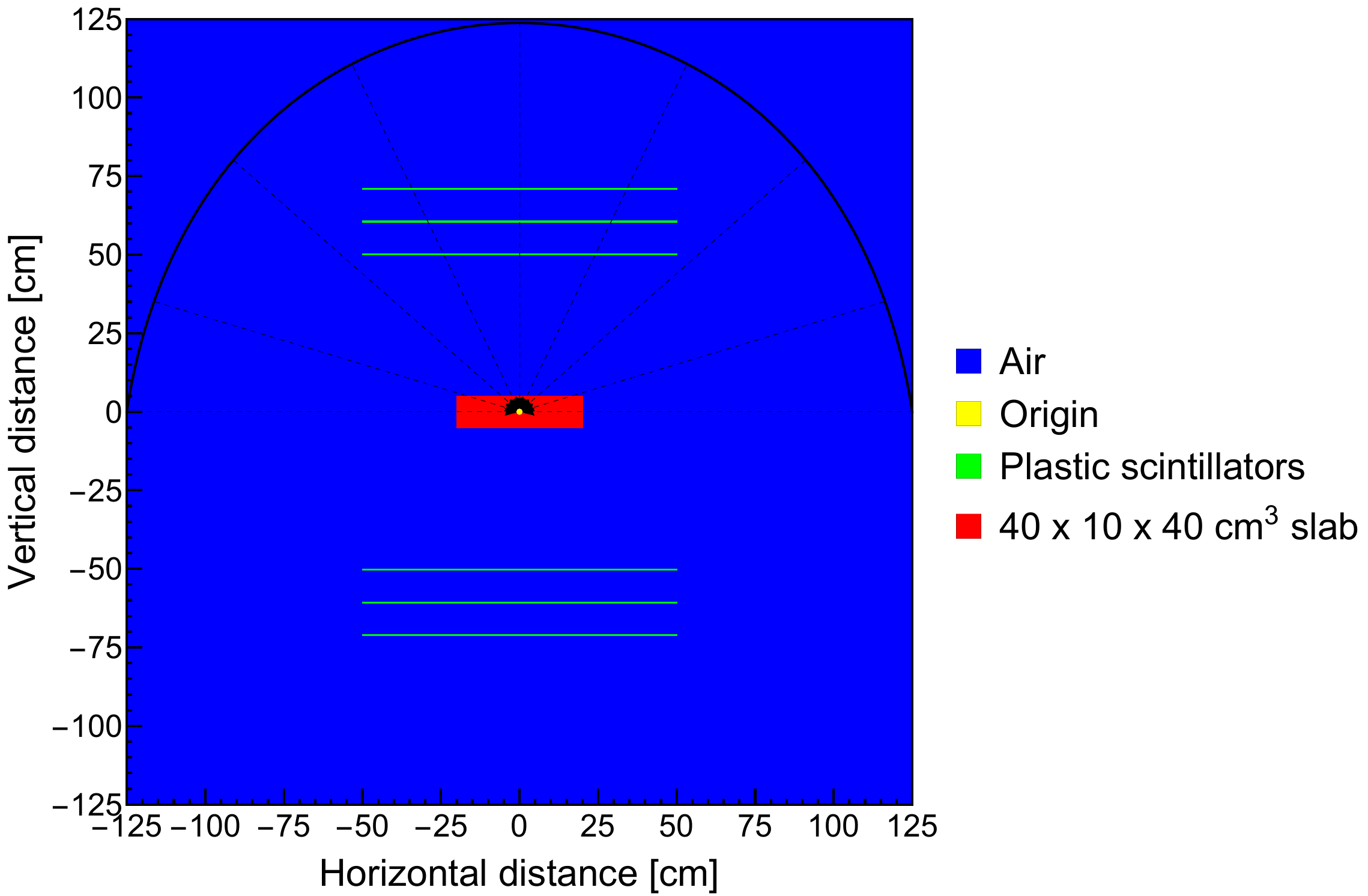}
\caption{Delineation of the generated particles from the hemispherical source with a momentum direction towards the origin.}
\end{center}
\label{Tomosetup}
\end{figure}
Our objective is to build a hemispherical muon source that surrounds our tomographic setup~\cite{georgadze2021method} similar to the other tomographic configurations existing in the literature~\cite{borozdin2003radiographic, frazao2019high, frazao2016discrimination} as delineated in Fig.~1. On the first basis, the particle locations in the Cartesian coordinates are generated by using the Gaussian distributions formally defined as G4RandGauss::shoot() in GEANT4 as written in
\begin{equation}
x_{0}=G(\bar{x},\sigma_{x},x)={\rm G4RandGauss::shoot()},
\end{equation}
and
\begin{equation}
y_{0}=G(\bar{y},\sigma_{y},y)={\rm G4RandGauss::shoot()},
\end{equation}
and
\begin{equation}
z_{0}=G(\bar{z},\sigma_{z},z)={\rm G4RandGauss::shoot()}.
\end{equation}
where $\bar{x}=\bar{y}=\bar{z}=0$ and $\sigma_{x}=\sigma_{y}=\sigma_{z}=1$ by definition. The generated spatial points are renormalized in order to form a unit sphere as indicated in
\begin{equation}
x_{0}^{*}=\frac{x_{0}}{\sqrt{x_{0}^{2}+y_{0}^{2}+z_{0}^{2}}},~~~
y_{0}^{*}=\frac{y_{0}}{\sqrt{x_{0}^{2}+y_{0}^{2}+z_{0}^{2}}},~~~
z_{0}^{*}=\frac{z_{0}}{\sqrt{x_{0}^{2}+y_{0}^{2}+z_{0}^{2}}}.
\end{equation}
Given a sphere of radius denoted by $R$,  the initial positions on the spherical surface of radius $R$ in cm in the Cartesian coordinates are obtained as follows
\begin{equation}
x_{i}=R*x_{0}^{*},~~~
y_{i}=R*|y_{0}^{*}|=R*{\rm ABS}(y_{0}^{*}),~~~
z_{i}=R*z_{0}^{*}.
\end{equation}
where the y-component of the Cartesian coordinates constituting the vertical axis is positively defined in order to yield the hemispherical surface. Then, the generated particles on the spherical surface are directed to the origin
\begin{equation}
x_{f}=0,~~~
y_{f}=0,~~~
z_{f}=0.
\end{equation}
Then, by constructing a vector from the hemispherical surface to the origin, one obtains
\begin{equation}
px=x_{f}-x_{i},~~~
py=y_{f}-y_{i},~~~
pz=z_{f}-z_{i}.
\end{equation}
Thus, the selective momentum direction denoted by $\vec{P}=(P_{x}, P_{y}, P_{z})$ is
\begin{equation}
P_{x}=\frac{px}{\sqrt{px^{2}+py^{2}+pz^{2}}},~~~
P_{y}=\frac{py}{\sqrt{px^{2}+py^{2}+pz^{2}}},~~~
P_{z}=\frac{pz}{\sqrt{px^{2}+py^{2}+pz^{2}}}.
\end{equation}
\subsection{Generation via coordinate transformation}
\begin{figure}[H]
\begin{center}
\includegraphics[width=9cm]{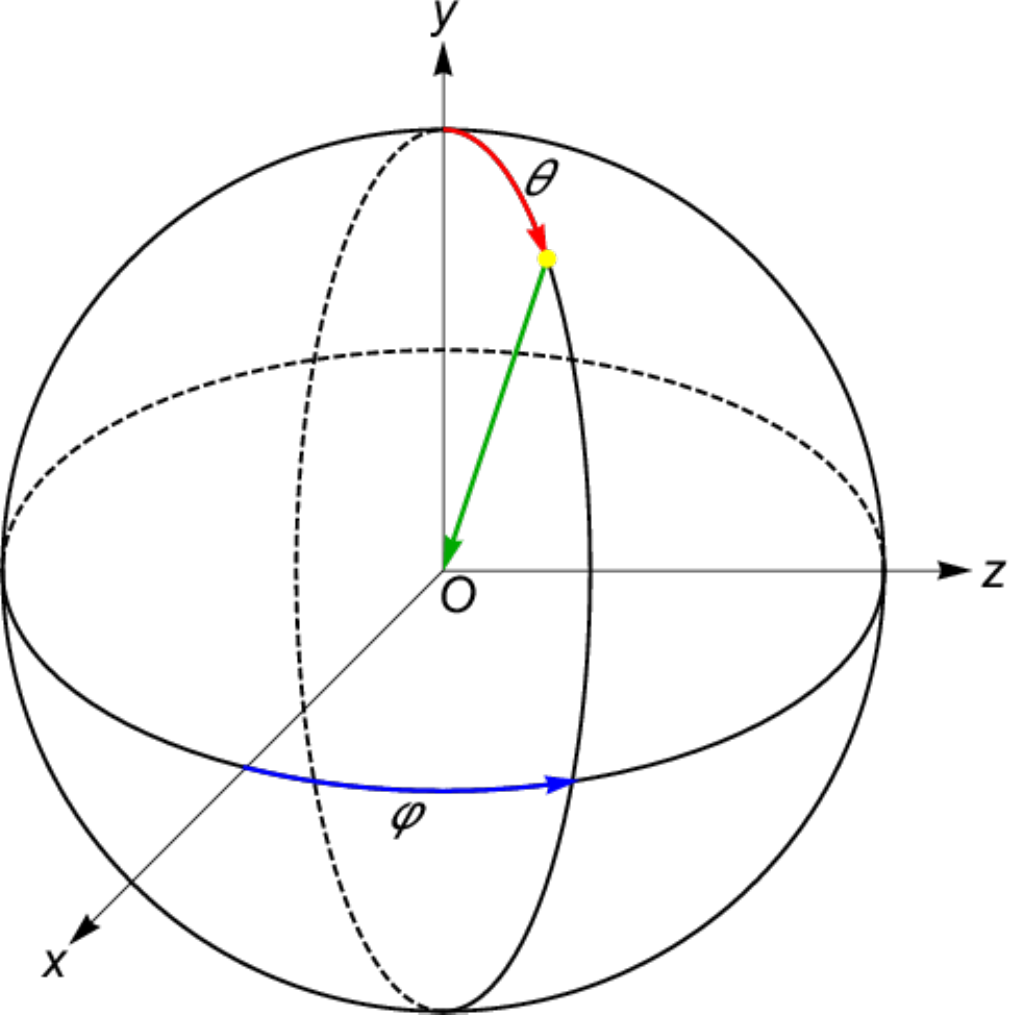}
\caption{Spherical variables consisting of $\theta$ and $\varphi$ with respect to the Cartesian coordinates (x,y,z).}
\end{center}
\end{figure}
The latter scheme is composed of the coordinate transformation as depicted in Fig.~2. To begin with, two numbers, i.e. $q_{1}$  and $q_{2}$, are uniformly generated to be inserted to the associated expression of the spherical variables as follows
\begin{equation}
q_{1}={\rm G4UniformRand()},
\end{equation}
and
\begin{equation}
q_{2}={\rm G4UniformRand()}.
\end{equation}
The surface generation is initiated by randomizing $\theta$ as well as $\varphi$ as shown in
\begin{equation}
\theta=\arccos{(2\times q_{1}-1)},
\end{equation} 
and
\begin{equation}
\varphi=2\times \pi\times q_{2} .
\end{equation}
The coordinate transformation yields the generated points on the hemispherical surface for a sphere of radius $R$ in the Cartesian coordinates as described in
\begin{equation}
x_{i}=R\times\sin\theta\times\cos\varphi
\end{equation}
and
\begin{equation} 
y_{i}=R\times|\cos\theta|=R\times{\rm ABS}(\cos\theta)
\end{equation}
and 
\begin{equation} 
z_{i}=R\times\sin\theta\times\sin\varphi
\end{equation}
where the y-component of the Cartesian coordinates constituting the vertical axis is repeatedly positively defined in order to yield the hemispherical surface as usual. Then, the generated particles on the spherical surface are again directed to the origin
\begin{equation}
x_{f}=0,~~~
y_{f}=0,~~~
z_{f}=0.
\end{equation}
Then, by constructing a vector from the hemispherical surface to the origin, one obtains
\begin{equation}
px=x_{f}-x_{i},~~~
py=y_{f}-y_{i},~~~
pz=z_{f}-z_{i}.
\end{equation}
Thus, the selective momentum direction denoted by $\vec{P}=(P_{x}, P_{y}, P_{z})$ is
\begin{equation}
P_{x}=\frac{px}{\sqrt{px^{2}+py^{2}+pz^{2}}},~~~
P_{y}=\frac{py}{\sqrt{px^{2}+py^{2}+pz^{2}}},~~~
P_{z}=\frac{pz}{\sqrt{px^{2}+py^{2}+pz^{2}}}.
\end{equation}
Finally, the simulation preview through either scheme is displayed in Fig.~3.
\begin{figure}[H]
\begin{center}
\includegraphics[width=9.5cm]{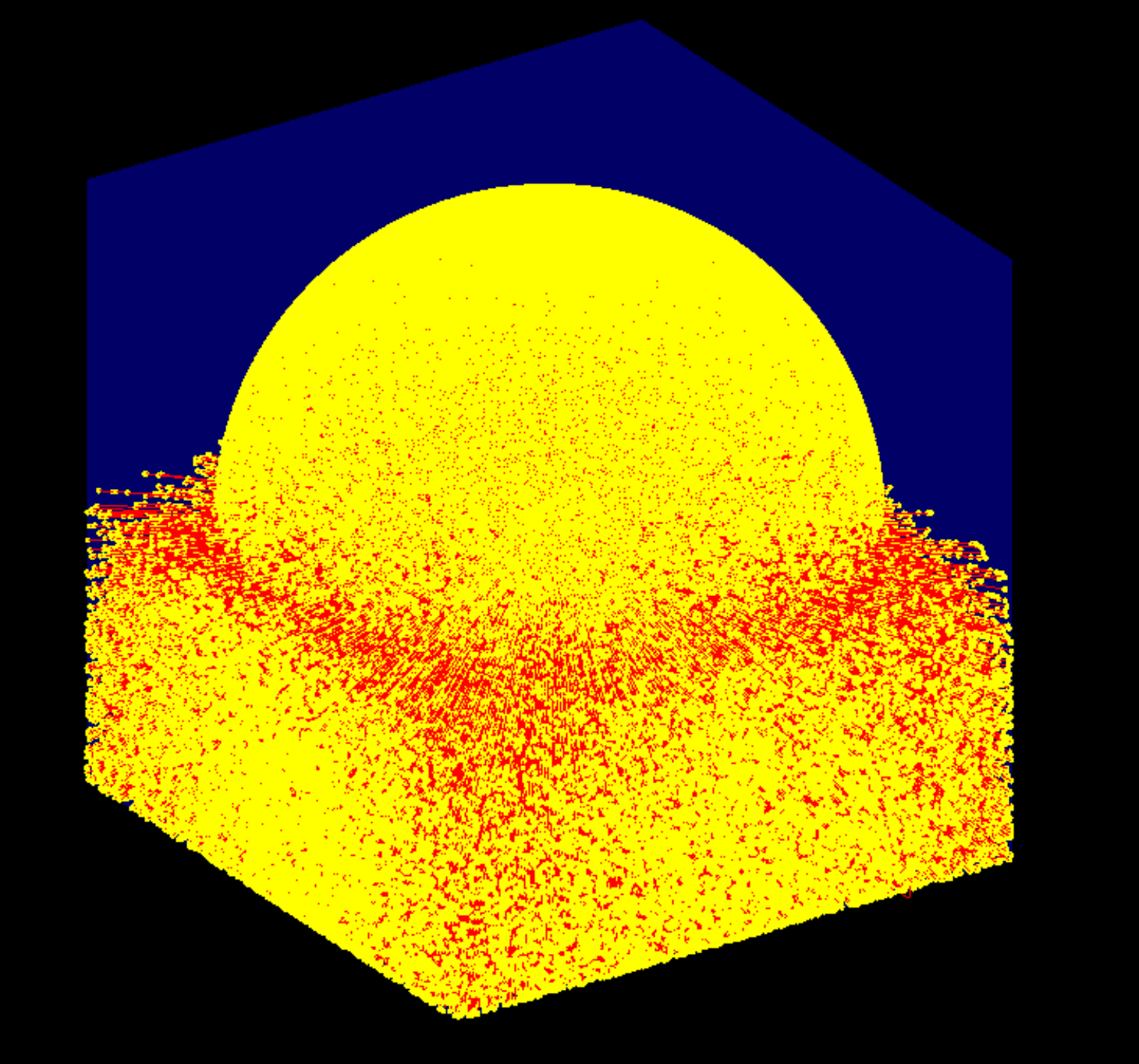}
\caption{Hemispherical muon source in GEANT4.}
\end{center}
\end{figure}
\section{Restrictive planar focus scheme}
As described in another study~\cite{topuz2022particle}, the generated particles from any initial point on the hemispherical surface might be directed to a location randomly selected on a pseudo plane that restricts the momentum direction, which also leads to the minimization of the particle escape. Thus, the particle locations in cm on a restrictive plane of $2L\times 2D$ $\rm cm^{2}$ situated at y=0 are supposed to have the spatial coordinates such that
\begin{equation}
x_{f}=-L+2\times L\times{\rm G4UniformRand()},~~~
y_{f}=0,~~~
z_{f}=-D+2\times D\times{\rm G4UniformRand()}.
\end{equation}
Then, by constructing a vector from the generative hemispherical surface to the restrictive plane, one obtains
\begin{equation}
px=x_{f}-x_{i},~~~
py=y_{f}-y_{i},~~~
pz=z_{f}-z_{i}.
\end{equation}
Thus, the selective momentum direction, i.e. $\vec{P}=(P_{x}, P_{y}, P_{z})$, is
\begin{equation}
P_{x}=\frac{px}{\sqrt{px^{2}+py^{2}+pz^{2}}},~~~
P_{y}=\frac{py}{\sqrt{px^{2}+py^{2}+pz^{2}}},~~~
P_{z}=\frac{pz}{\sqrt{px^{2}+py^{2}+pz^{2}}}.
\end{equation}
\section{Conclusion}
In this study, we explore the possibility to use the random number generators that are already defined in the GEANT4 code. By profiting from these random number generators, we provide a number of source schemes where the first strategy is based on the Gauss distributions, whereas the latter procedure requires a coordinate transformation by utilizing the spherical variables. Finally, we obtain a hemispherical muon source where the kinetic energies of the generated muons are discretized, and the momentum directions of these generated muons are selective by means of the vector constructions. We call this source discrete oriented muon emission (DOME). DOME has been developed for simulations of muon tomography scenarios where the volume of interest is contained in a gap between detection layers, and the hemispheric source surrounds the whole setup. However, it can find applications in a broader array of use cases. For example, as demonstrated in~\cite{pagano2021ecomug}, hemispheric sources are computationally efficient and at the same time unbiased for measurements of the cosmic muon flux where the detector has complex geometry. Moreover, nothing prevents applications of the same method in simulations of muon radiography setups for volcanoes or pyramids or other very large objects of interest that are distant from the detector~\cite{bonechi2020atmospheric}, and solid angle restrictions can optionally be imposed to increase computational efficiency.
\section*{Appendix A - Generation via Gaussian distributions }
\begin{tiny}
\begin{lstlisting}
#include "B1PrimaryGeneratorAction.hh"
#include "G4LogicalVolumeStore.hh"
#include "G4LogicalVolume.hh"
#include "G4Box.hh"
#include "G4RunManager.hh"
#include "G4ParticleGun.hh"
#include "G4ParticleTable.hh"
#include "G4ParticleDefinition.hh"
#include "G4SystemOfUnits.hh"
#include "Randomize.hh"
#include <iostream>
using namespace std;

B1PrimaryGeneratorAction::B1PrimaryGeneratorAction()
: G4VUserPrimaryGeneratorAction(),
  fParticleGun(0)
//  fEnvelopeBox(0)
{
  G4int n_particle = 1;
  fParticleGun  = new G4ParticleGun(n_particle);

  // default particle kinematic
  G4ParticleTable* particleTable = G4ParticleTable::GetParticleTable();
  G4String particleName;
  G4ParticleDefinition* particle
    = particleTable->FindParticle(particleName="mu-");
  fParticleGun->SetParticleDefinition(particle);
}

B1PrimaryGeneratorAction::~B1PrimaryGeneratorAction()
{
  delete fParticleGun;
}

//80-bin Discrete CRY Energy Spectrum
void B1PrimaryGeneratorAction::GeneratePrimaries(G4Event* anEvent)
{
//Discrete probabilities
double A[]= {0.0, 0.01253639, 0.02574546, 0.02802035, 0.02706636, 0.03528534, 0.02826496,
0.03157946, 0.03078447, 0.02777574, 0.02546415, 0.03150608, 0.02815489,
0.02580661, 0.02364179, 0.02170935, 0.02152589, 0.02348279, 0.02134243,
0.0196913,  0.02036398, 0.01841931, 0.01718402, 0.01700056, 0.01624226,
0.01539835, 0.01536166, 0.01471344, 0.01422421, 0.01412637, 0.01284215,
0.01260977, 0.01213278, 0.0129033,  0.01248746, 0.01196155, 0.01064064,
0.01057949, 0.0096255,  0.0103838,  0.00928304, 0.00879382, 0.00884274,
0.00793767, 0.00786429, 0.00769306, 0.00709376, 0.00736283, 0.0071916,
0.00721607, 0.00692253, 0.00643331, 0.00678799, 0.00673907, 0.00618869,
0.00634769, 0.00665346, 0.00650669, 0.00561385, 0.00589516, 0.00589516,
0.00578508, 0.00557716, 0.00550378, 0.00434187, 0.0043541,  0.00408503,
0.00364472, 0.00399941, 0.00388934, 0.00396272, 0.00431741, 0.00368142,
0.00363249, 0.00362026, 0.00410949, 0.00336342, 0.00358357, 0.00362026,
0.00348573, 0.0035958}; 
//Discrete energies
double B[]= {0.0, 100, 200, 300, 400, 500, 600, 700, 800, 900, 1000, 
  1100, 1200, 1300, 1400, 1500, 1600, 1700, 1800, 1900, 2000, 
  2100, 2200, 2300, 2400, 2500, 2600, 2700, 2800, 2900, 3000, 
  3100, 3200, 3300, 3400, 3500, 3600, 3700, 3800, 3900, 4000, 
  4100, 4200, 4300, 4400, 4500, 4600, 4700, 4800, 4900, 5000, 
  5100, 5200, 5300, 5400, 5500, 5600, 5700, 5800, 5900, 6000, 
  6100, 6200, 6300, 6400, 6500, 6600, 6700, 6800, 6900, 7000, 
  7100, 7200, 7300, 7400, 7500, 7600, 7700, 7800, 7900, 8000};
G4int SizeEnergy=sizeof(B)/sizeof(B[0]);
G4int SizeProbability=sizeof(A)/sizeof(A[0]);

G4double Grid[sizeof(B)/sizeof(B[0])];
double sum=0;
  for(int x=0; x < 81; x++){
  sum=sum+A[x];
  Grid[x]=sum;
  std::ofstream GridFile;
  GridFile.open("Probability_grid.txt", std::ios::app);
  GridFile <<  Grid[x] << G4endl;
  GridFile.close();
  }
  G4double radius=100*cm; //radius of sphere
  for (int n_particle = 1; n_particle < 100000; n_particle++){
  G4double x0=G4RandGauss::shoot();
  std::ofstream GaussFile;
  GaussFile.open("Gauss_x.txt", std::ios::app); //in mm
  GaussFile << x0 << G4endl;
  GaussFile.close();
//Centerally focused semi-spherical source via Gauss distributions
  G4double y0=G4RandGauss::shoot();
  G4double z0=G4RandGauss::shoot();
  G4double n0=sqrt(pow(x0,2)+pow(y0,2)+pow(z0,2));
//Coordinates on sphere  
  x0 = radius*(x0/n0); 
  y0 = radius*abs(y0/n0);
  z0 = radius*(z0/n0);
  std::ofstream SphereFile;
  SphereFile.open("coordinates_on_sphere.txt", std::ios::app); //in mm
  SphereFile << x0 << " "<< y0 << " " << z0 << " " << G4endl;
  SphereFile.close();   
  fParticleGun->SetParticlePosition(G4ThreeVector(x0,y0,z0));
//Aimed at origin
  G4double x1=0;
  G4double y1=0;
  G4double z1=0;
  G4double mx = x1-x0;
  G4double my = y1-y0;
  G4double mz = z1-z0;
  G4double mn = sqrt(pow(mx,2)+pow(my,2)+pow(mz,2));
  mx = mx/mn;
  my = my/mn;
  mz = mz/mn;    
  fParticleGun->SetParticleMomentumDirection(G4ThreeVector(mx,my,mz));
  G4double Energy=0; //Just for initialization
  G4double pseudo=G4UniformRand();
  for (int i=0; i < 81; i++){
  if(pseudo > Grid[i] && pseudo <= Grid[i+1]){
  Energy=B[i+1];
  std::ofstream EnergyFile;
  EnergyFile.open("Energy.txt", std::ios::app);
  EnergyFile <<  Energy << G4endl;
  EnergyFile.close();
  } 
  }   
  fParticleGun->SetParticleEnergy(Energy);
  fParticleGun->GeneratePrimaryVertex(anEvent);
  }
  }
\end{lstlisting}
\end{tiny}
\section*{Appendix B - Generation by means of coordinate transformation}
\begin{tiny}
\begin{lstlisting}
#include "B1PrimaryGeneratorAction.hh"
#include "G4LogicalVolumeStore.hh"
#include "G4LogicalVolume.hh"
#include "G4Box.hh"
#include "G4RunManager.hh"
#include "G4ParticleGun.hh"
#include "G4ParticleTable.hh"
#include "G4ParticleDefinition.hh"
#include "G4SystemOfUnits.hh"
#include "Randomize.hh"
#include <iostream>
using namespace std;

B1PrimaryGeneratorAction::B1PrimaryGeneratorAction()
: G4VUserPrimaryGeneratorAction(),
  fParticleGun(0)
//  fEnvelopeBox(0)
{
  G4int n_particle = 1;
  fParticleGun  = new G4ParticleGun(n_particle);

  // default particle kinematic
  G4ParticleTable* particleTable = G4ParticleTable::GetParticleTable();
  G4String particleName;
  G4ParticleDefinition* particle
    = particleTable->FindParticle(particleName="mu-");
  fParticleGun->SetParticleDefinition(particle);
}

B1PrimaryGeneratorAction::~B1PrimaryGeneratorAction()
{
  delete fParticleGun;
}

//80-bin Discrete CRY Energy Spectrum
void B1PrimaryGeneratorAction::GeneratePrimaries(G4Event* anEvent)
{
//Discrete probabilities
double A[]= {0.0, 0.01253639, 0.02574546, 0.02802035, 0.02706636, 0.03528534, 0.02826496,
0.03157946, 0.03078447, 0.02777574, 0.02546415, 0.03150608, 0.02815489,
0.02580661, 0.02364179, 0.02170935, 0.02152589, 0.02348279, 0.02134243,
0.0196913,  0.02036398, 0.01841931, 0.01718402, 0.01700056, 0.01624226,
0.01539835, 0.01536166, 0.01471344, 0.01422421, 0.01412637, 0.01284215,
0.01260977, 0.01213278, 0.0129033,  0.01248746, 0.01196155, 0.01064064,
0.01057949, 0.0096255,  0.0103838,  0.00928304, 0.00879382, 0.00884274,
0.00793767, 0.00786429, 0.00769306, 0.00709376, 0.00736283, 0.0071916,
0.00721607, 0.00692253, 0.00643331, 0.00678799, 0.00673907, 0.00618869,
0.00634769, 0.00665346, 0.00650669, 0.00561385, 0.00589516, 0.00589516,
0.00578508, 0.00557716, 0.00550378, 0.00434187, 0.0043541,  0.00408503,
0.00364472, 0.00399941, 0.00388934, 0.00396272, 0.00431741, 0.00368142,
0.00363249, 0.00362026, 0.00410949, 0.00336342, 0.00358357, 0.00362026,
0.00348573, 0.0035958}; 
//Discrete energies
double B[]= {0.0, 100, 200, 300, 400, 500, 600, 700, 800, 900, 1000, 
  1100, 1200, 1300, 1400, 1500, 1600, 1700, 1800, 1900, 2000, 
  2100, 2200, 2300, 2400, 2500, 2600, 2700, 2800, 2900, 3000, 
  3100, 3200, 3300, 3400, 3500, 3600, 3700, 3800, 3900, 4000, 
  4100, 4200, 4300, 4400, 4500, 4600, 4700, 4800, 4900, 5000, 
  5100, 5200, 5300, 5400, 5500, 5600, 5700, 5800, 5900, 6000, 
  6100, 6200, 6300, 6400, 6500, 6600, 6700, 6800, 6900, 7000, 
  7100, 7200, 7300, 7400, 7500, 7600, 7700, 7800, 7900, 8000};
G4int SizeEnergy=sizeof(B)/sizeof(B[0]);
G4int SizeProbability=sizeof(A)/sizeof(A[0]);

G4double Grid[sizeof(B)/sizeof(B[0])];
double sum=0;
  for(int x=0; x < 81; x++){
  sum=sum+A[x];
  Grid[x]=sum;
  std::ofstream GridFile;
  GridFile.open("Probability_grid.txt", std::ios::app);
  GridFile <<  Grid[x] << G4endl;
  GridFile.close();
  }
  G4double radius=100*cm; //radius of sphere
  for (int n_particle = 1; n_particle < 100000; n_particle++){
//Centerally focused semi-spherical source via coordinate transformation
  G4double rand1=G4UniformRand();
  G4double rand2=G4UniformRand();
  G4double theta=acos(2*rand1-1);
  G4double phi=2*3.14159265359*rand2;
//Coordinates on sphere 
  G4double x0=radius*sin(theta)*cos(phi);
  G4double y0=radius*abs(cos(theta));
  G4double z0=radius*sin(theta)*sin(phi);
  std::ofstream SphereFile;
  SphereFile.open("coordinates_on_sphere.dat", std::ios::app); //in mm
  SphereFile << x0 << " "<< y0 << " " << z0 << G4endl;
  SphereFile.close();   
  fParticleGun->SetParticlePosition(G4ThreeVector(x0,y0,z0));
//Aimed at origin
  G4double x1=0;
  G4double y1=0;
  G4double z1=0;
  G4double mx = x1-x0;
  G4double my = y1-y0;
  G4double mz = z1-z0;
  G4double mn = sqrt(pow(mx,2)+pow(my,2)+pow(mz,2));
  mx = mx/mn;
  my = my/mn;
  mz = mz/mn;    
  fParticleGun->SetParticleMomentumDirection(G4ThreeVector(mx,my,mz));
  G4double Energy=0; //Just for initialization
  G4double pseudo=G4UniformRand();
  for (int i=0; i < 81; i++){
  if(pseudo > Grid[i] && pseudo <= Grid[i+1]){
  Energy=B[i+1];
  std::ofstream EnergyFile;
  EnergyFile.open("Energy.txt", std::ios::app);
  EnergyFile <<  Energy << G4endl;
  EnergyFile.close();
  } 
  }   
  fParticleGun->SetParticleEnergy(Energy);
  fParticleGun->GeneratePrimaryVertex(anEvent);
  }
  }
\end{lstlisting}
\end{tiny}
\bibliographystyle{ieeetr}
\bibliography{DOMEbiblio} 
\end{document}